  \documentstyle[12pt,amssymbols]{article}
  \input math_macros.tex

\begin{document}
  \begin{titlepage}
  \begin{center}
  \today     \hfill    LBL-37944 \\

  \vskip .4in

  {\large \bf Chance, Choice, and Consciousness:\\
  The Role of Mind in the Quantum Brain}
  \footnote{This work was supported by the Director, Office of Energy
  Research, Office of High Energy and Nuclear Physics, Division of High
  Energy Physics of the U.S. Department of Energy under Contract
  DE-AC03-76SF00098.}
  \vskip .40in

  Henry P. Stapp\\
  {\em Lawrence Berkeley Laboratory\\
        University of California\\
      Berkeley, California 94720}
  \end{center}

  \vskip .5in

  \begin{abstract}
   Contemporary quantum mechanical description of nature involves
  two processes. The  first is a  dynamical process governed  by the equations
of
  local quantum  field theory.  This process is  local and  deterministic, but
it
  generates a  structure that is not  compatible with  observed reality. A
second
  process  is  therefore   invoked. This  second   process  somehow  analyzes
the
  structure   generated  by the  first  process  into a   collection of
possible
  observable   realities, and  selects  one of  these as the  actually
appearing
  reality.  This  selection  process is not  well  understood. It is
necessarily
  nonlocal and,  according to  orthodox  thinking, is governed  by an
irreducible
  element of chance. The  occurrence of this  irreducible element of chance
means
  that the  theory is not  naturalistic:  the  dynamics is  controlled in part
by
  something that is not part of the physical universe. The present work
describes
  a quantum  mechanical model of  brain dynamics  in which the  quantum
selection
  process  is a  causal  process  governed  not by  pure  chance but  rather by
a
  mathematically    specified  nonlocal   physical  process   identifiable as
the
  conscious process.  \end{abstract}

  \end{titlepage}
  \renewcommand{\thepage}{\roman{page}}
  \setcounter{page}{2}
  \mbox{ }

  \vskip 1in

  \begin{center}
  {\bf Disclaimer}
  \end{center}

  \vskip .2in

  \begin{scriptsize}
  \begin{quotation}
  This document was prepared as an account of work sponsored by the United
  States Government. While this document is believed to contain correct
   information, neither the United States Government nor any agency
  thereof, nor The Regents of the University of California, nor any of their
  employees, makes any warranty, express or implied, or assumes any legal
  liability or responsibility for the accuracy, completeness, or usefulness
  of any information, apparatus, product, or process disclosed, or represents
  that its use would not infringe privately owned rights.  Reference herein
  to any specific commercial products process, or service by its trade name,
  trademark, manufacturer, or otherwise, does not necessarily constitute or
  imply its endorsement, recommendation, or favoring by the United States
  Government or any agency thereof, or The Regents of the University of
  California.  The views and opinions of authors expressed herein do not
  necessarily state or reflect those of the United States Government or any
  agency thereof or The Regents of the University of California and shall
  not be used for advertising or product endorsement purposes.
  \end{quotation}
  \end{scriptsize}

  \vskip 2in

  \begin{center}
  \begin{small}
  {\it Lawrence Berkeley Laboratory is an equal opportunity employer.}
  \end{small}
  \end{center}

  \newpage
  \renewcommand{\thepage}{\arabic{page}}
  \setcounter{page}{1}
  \noindent {\bf 1. Introduction.}

  The orthodox  Copenhagen  interpretation of  quantum theory,  as promulgated
by
  Niels Bohr, is pragmatic: the quantum  formalism is regarded as merely a set
of
  useful rules for  predicting what our classically  describable experiences
will
  be under conditions specified by classically describable experiences. The
entry
  of chance into the theory  is regarded as an  expression of the brute
empirical
  fact that atomic systems prepared and measured identically---to the best of
our
  experimental  capabilities--- give  nonidentical results  that have
statistical
  regularities.  The  question of  the origin  of the  element of  chance is
left
  unanswered. Einstein  rejected the idea that God  plays dice with the
universe,
  and Bohr concurred in rejecting the  idea of ``a choice on the part of
`nature'
  ''[1]. Yet the notion that a definite choice is fixed by nothing at all is
even
  more repugnant to rational thought.

  Bohr's interpretation  does not cover biological  and cosmological systems.
The
  possibility therefore arises that what appears as pure chance in the
restricted
  domain of  atomic  phenomena has its  roots in a  more complete  description
of
  nature.

  The element of  chance normally enters  into quantum  theory in connection
with
  our observations. In the absence of  observations the evolution of the
universe
  is  governed by local  laws that  are  natural  generalizations of  the laws
of
  classical mechanics: the universe is  conceived to be an aggregate of
localized
  properties,   and  the  rate of   change in  each  such   property is
governed
  exclusively by nearby properties.  Observations, however, are associated with
a
  ``second process''  that is logically  required to be  highly nonlocal [2],
and
  therefore fundamentally different from the first process.

  The aim of the present work is to  provide a logical and mathematical
framework
  for a  causal theory  of brain  dynamics in  which the   controlling element
of
  chance in  the quantum  selection  process is  replaced by a  nonlocal
physical
  process identifiable with the conscious aspect of brain process.

  In this formulation of quantum  dynamics conscious experiences exercise
genuine
  control over brain activity. Analogous  elements should occur in all
biological
  systems, due to the  enormous survival advantage  they can confer. But in
lower
  life forms,  and also in  the  inanimate part of  nature, these  elements
will,
  because of the  absence in  them of the  intentional and  attentional
structure
  supplied by our brains, be very different from human conscious experiences.

  \noindent {\bf 2. Quantum Searching and Survival.}

  Survival,  at  least in the  animal  kingdom,  depends on  rapidly  finding
and
  executing   appropriate  behaviors. Options  are  generally  available, and
the
  organism must reject  those not appropriate in the  specific situation in
which
  it finds itself, and  pursue one that is  appropriate. The process of
searching
  for an  appropriate behavior  can be likened  to a search for  the way out of
a
  maze. The  classical  search procedure  is  essentially to try,  at some
mental
  level, each of the  possibilities until a blockage  is encountered, and then
to
  back off and try another. This can be very time consuming, and an organism
that
  uses it is likely to be devoured by  one that employs a faster process.
Massive
  parallel   (and   interconnected)   processing  may  offer   advantages, but
it
  introduces the compensating problem of  keeping the whole system operating in
a
  coherently coordinated way.

  For rapid  searching the  exploitation of the  quantum  character of brains
can
  confer a huge advantage. Quantum  dynamics is essentially hydrodynamics [3].
The
  contrast between classical and quantum  search procedures can be likened to
the
  contrast between  the particle and  hydrodynamical  solutions to the problem
of
  getting out of a maze: in  the particle solution  the particle bounces
randomly
  around the maze in the hope of finding the small opening; in the
hydrodynamical
  solution  the maze is  filled  with water,  which then  rushes out  through
the
  opening. The  essential point is that  in  classical-particle dynamics what
the
  particle does  is completely  unaffected by  what it would  have done if it
had
  been on a nearby  trajectory, whereas the flow of  water is affected by what
is
  happening nearby: if water rushes out at one place, leaving a void, then
nearby
  water rushes in to fill the void, sucking in water from further away.

  This point can be illustrated by  considering a circular trough that has also
a
  circular cross section in each radial plane. Suppose this trough is filled
with
  a statistical ensemble representing  alternative possibilities for the
position
  and velocity  of one  particle. Each  element of  the ensemble  oscillates in
a
  radial plane, with no  angular motion. Suppose we  open a small angular
section
  of the trough  so that the  particles in that  section flow  out. The
remaining
  particles,  which  represent the  alternative  possibilities,  will continue
to
  oscillate forever. But if one fills the trough with water and opens the
section
  then all the  water runs  out. The  quantum  probability  function for one
  particle behaves like water, not like the statistical ensemble of independent
  particles.

  A physicist who wants to see this in the equations can consider a wave
function
  for a particle confined to a circle. The time-dependent Schroedinger has on
the
  left the operator  i times the  derivative with respect to  t, and on the
right
  the  kinetic  energy term.  To  represent the  opening in  the maze  (i.e.,
the
  solution that is not blocked by negative feed-back) add on the right the term
b
  times minus i times a  Dirac delta function of the  (cyclic) argument x(mod
1).
  Then the rate of loss of probability  in the ring is 2b times the square of
the
  magnitude of the wave function at x=0.  This is non-negative, and more
detailed
  calculations  show that  the  probability is  rapidly sucked to  the point
x=0,
  where it disappears.

  A more  realistic  model  would have  in place  of the  Dirac delta  function
a
  function with a flat  central plateau bounded on  each side by a sharp
gaussian
  fall-off. The  rate of flow of probability from the surrounding region into
the
  region of probability loss is controlled by the sharpness of the gaussian
walls.

  This way of searching for an  appropriate response should be particularly
rapid
  and effective in a brain organized in the way described in [4], because in
that
  system the unblocked flow  out of the maze (of  alternatives, most of which
are
  blocked  by  negative  feed-back)  creates a  template for  action,  which
then
  automatically evolves into the corresponding action itself. There is no need
to
  convert the solution  represented by the unblocked  flow into a plan of
action,
  and then to create the corresponding  sets of instructions to muscles etc.:
the
  unimpeded flow produces a template for action that, if not blocked,
automatically evolves
  into the appropriate  action itself. So the basic  problem of rapidly
producing
  an   appropriate  action is   precisely  that of   rapidly  getting  all of
the
  probability into an unblocked channel, i.e., of keeping the search process
from
  getting hung up exploring the blocked channels. The hydrodynamical character
of
  the quantum law of evolution provides an efficient way to solve this problem.

  Notice  also that  the  quantum  mechanism does  not  involve a  sudden `all
or
  nothing' leap in phylogenetic  development: even a little bit of sucking of
the
  probabilities into  unblocked channels will aid  survival, and the organism
can
  gradually evolve in a way that tends to enhance the process.

  \noindent {\bf 3. Decoherence}

  It has often been observed that the coupling of a system to its environment
has
  a tendency to make interference  phenomena that are present in principle
within
  quantum systems  difficult to observe  in practice. Phase  relationships,
which
  are essential to interference phenomena, get diffused into the environment,
and
  are difficult to  retrieve. The net  effect of this is to  make a large part
of
  the  observable   phenomena in a  quantum  universe  similar  to what  would
be
  observed in a world in  which certain collective  (i.e., macroscopic)
variables
  are  governed by  classical  mechanics. This  greatly  diminishes  the realm
of
  phenomena  that require  for their  understanding  the explicit  use of
quantum
  theory.

  These decoherence effects will have a tendency to reduce, in a system such as
  the brain, the distances over which the idea of a simple single quantum
system
  holds. This will reduce the distances over which the simple hydrodynamical
  considerations described above will hold. However, the following points
  must be considered.

  a) A calcium ion entering a bouton through a microchannel of diameter $x$
must,
  by Heisenberg's  indeterminacy principle, have a  momentum spread of
$\hbar/x$,
  and hence a  velocity spread  of  $(\hbar/x)/m$, and hence a  spatial spread
in
  time $t$, if the particle were freely  moving, of $t(\hbar/x)/m$. Taking $t$
to
  be  $200$  microseconds,  the  typical  time for  the ion to  diffuse  from
the
  microchannel  opening  to a  triggering site  for the  release of  a vesicle
of
  neurotransmitter, and taking $x$ to be one nanometer, one finds the diameter
of
  the wave function to be about $0.04$ centimeters, which is huge compared to
the
  $1/100000000$ centimeter size of the calcium ion. There is, therefore, in
brain
  dynamics  a  powerful  counterforce to  the  mechanisms  that tend  to
diminish
  quantum coherence effects.

  b) The  normal process  that induces  decoherence  arises from  the fact that
a
  collision  of  a state   represented  by a  broad wave   function with  a
state
  represented by a narrow wave packet effectively reduces the coherence length
in
  the first state to  a distance  proportional to the width  of the second
state.
  But in an aqueous medium in which all the states of the individual systems
have
  broad  packets this  mechanism is no  longer  effective:  coherence lengths
can
  remain long.

  c) Even if the coherence length were only a factor of ten times the diameter
of
  the atom or ion involved in some process, the cross section involved would be
a
  hundred times  larger. The search  processes under  consideration here
involves
  huge numbers of atoms and  ions acting together,  and the cross-section
factors
  multiply. Thus  even a small  effect at the  level of the  individual atoms
and
  ions  could  give, by  virtue  of the hydrodynamical  effect,  a large
quantum
  enhancement  of the  efficiency of an  essentially  aqueous  macroscopic
search
  process.

  \noindent {\bf 4. Quantum Theory and Experience}

  The core problem  in quantum theory is  perhaps best  illustrated by
Einstein's
  example [5] of a radioactive atom  placed in a Geiger counter that is hooked
up
  to a pen  that is  drawing a  line on a  moving strip  of paper:  when the
atom
  decays the  Geiger  counter fires, and  this causes  a blip to  be drawn on
the
  moving strip  of paper.  Since all the  parts of  the apparatus  are made up
of
  atoms and electrons,  etc., one should be able to  apply quantum theory. But
if
  one simply applies the  Schroedinger equation, or  the equations of local
field
  theory,  one  finds  that  the  moving  strip  will  evolve  into a
continuous
  superposition of  possibilities, with every  possible time of decay
represented
  by a  correspondingly placed blip. No  single decay time  is singled out as
the
  actual decay  time. But what  is observed if  one looks at the  strip is a
blip
  appearing in  one place,  rather than  a smeared  out  superposition of all
the
  possibilities.   So  quantum  theory,  if left  in  this stage  where  only
the
  Schroedinger  equation (or the  corresponding  equation of  local quantum
field
  theory) is considered, is incomplete:  some explanation of the mismatch
between
  what we   experience and  what is  generated by  the  Schroedinger  equation
is
  needed. Some account is needed for the process that selects, from the
continuum
  of possibilities generated by Schroedinger equation , the particular thing
that
  we actually see.

  Physicists have proposed a number of  possible ways of completing the theory.
I
  do not wish to describe  them here in detail. The  chief contenders can be
tied
  to the names of Bohr, Bohm, Everett,  Heisenberg, and Wigner. Very briefly,
the
  essence of each position is as follows:

  \noindent   {\bf  Bohr [6]:}   Quantum  theory is  a set  of  useful  rules
that
  scientists  can use  to compute  statistical   prediction about  whether or
not
  certain  conceivable   classically  describable  experiences will  appear
under
  various conditions specified by classical describable experiences.  Defect:
The
  theory   formulated in  this  way  admittedly  does  not cover   biological
and
  cosmological   systems,  hence a  putative  theoretical  description  of
nature
  herself might be useful for the further development of science.

  \noindent {\bf Bohm [7]:} There is in addition to the quantum wave function
also
  a real classical world  whose motion is controlled  by the wave function. As
in
  classical mechanics the entire course  of history is fixed at the moment of
the
  creation of the universe. Defects:  This formulation is very nonparsimonious
  because the  Schroedinger  equation must grind  out forever  the infinitudes
of
  ``empty branches'' of the wave  function that will never have any effect on
the
  the  classical world,  which is the  only part of  reality that  we
experience.
  Also, the statistical  aspects of quantum theory  enter though the obscure
idea
  of a preferred  statistical ensemble  of universes.  Finally, consciousness
can
  play no causal role in the dynamics.

  \noindent {\bf Everett  [8]:} The wave function of  the universe is
continually
  separating into  ``branches'' that are  ``decoherent'' in the sense that if
one
  restricts the set  observables to certain  localizable collective
(macroscopic)
  properties  then the  state of the  universe is  approximately  equivalent to
a
  statistical  mixture of these  branches. It is assumed  that there are
separate
  mental states  associated with  these separate  branches, and  that they can
be
  treated as  members of a  statistical  ensemble  with weights  specified by
the
  weights of  the  corresponding  statistical  ensemble  of  branches. All of
the
  mental states in  this ensemble are  assumed to really  exist, even though
each
  such state  contains no awareness of  the others. Defects:  This formulation
is
  very  nonparsimonious: only one of the  infinitude of  mental universes is
ever
  experienced by  us. Also, the  treatment of  the mental states  does not
follow
  from  the  physics: the  state of  the  universe is  a   ``conjunction'' of
the
  branches (it  consists of  branch 1 and branch  2 and ...)  whereas in order
to
  apply statistics the set of mental  worlds must be ``disjunctive'' (it
consists
  of mind 1 or mind 2 or  ...). The notion that a  single mental state can
evolve
  into either  mental state 1 or  mental state  2, with  specified
probabilities,
  seems incompatible with  the idea that the two  alternatives are
simultaneously
  present  and  really  existing. At the  very  least, these  ideas  constitute
a
  radical departure from normal ideas about the relationship between
conjunctions
  and disjunctions. Furthermore, the notion that the wave function separates
into
  well  defined  distinct   ``branches'' is  not  always  applicable:  the
normal
  evolution  of a wave  is an  amorphous  spreading out.  This  creates a
serious
  technical problem, not  yet resolved, of how to  define the decomposition of
an
  amorphous  quantum  structure  into a  disjunction of  classically
describable
  observable   realities in  such a  way  that a  probability  can be
coherently
  assigned to each of the  associated overlapping  mind/brain states, if there
is
  no physical  process that  picks out  and  actualizes one of  these
overlapping
  states, and rejects the others.  Finally, consciousness can play no causal
role
  in the dynamics.

  \noindent {\bf  Heisenberg [9]:}  Heisenberg is a  co-creator of the
Copenhagen
  interpretation that I have associated with Bohr. But he also proposed a
picture
  of nature herself in which there are  to kinds of realities: potentialities
and
  actualities. It is possible to regard  the wave function as a representation
of
  ``potentialities'' for  ``actual events'': the  potentialities evolve
according
  to the  Schroedinger equation until  the conditions for a  possible `event'
are
  created, and then  this event either  occurs or does not  occur, according to
a
  prescribed statistical rule. If the event occurs then the wave function
changes
  to a new form that reflects the fact  that this event has occurred, and then
it
  (the new wave function) proceeds again  to evolve according to the
Schroedinger
  equation.  The events  are  supposed to  occur in  connection with
``measuring
  devices''. Defects: The definition of ``measuring device'' is not specified,
  and hence the theory is not well defined. And, again, mind plays no role in
the
  dynamics

  \noindent {\bf Wigner [10]:} Wigner, giving credit to von Neumann, suggests
that
  what     characterizes a    ``measuring   device''  is  the   occurrence  of
an
  ``experience''    in   connection with  the   measurement.    Specifically,
the
  ``measuring devices'' of the  Heisenberg interpretation are identified with
the
  aspects  of  brain  dynamics  directly   associated  with the   occurrence of
a
  conscious experience.

  Each of  these general  approaches has  its  contemporary  proponents. Thus
the
  works of Ghirardi, Rimini, Weber, and Pearle [11] are in the Heisenberg
spirit.
  The works of Gell-mann and Hartle [12]  are in the Everett framework. The
works
  of Omnes [13] are,  apparently, in the Bohr spirit.  The present work is in
the
  Heisenberg-Wigner-von  Neumann spirit: I accept the  general idea of
Heisenberg
  that the wave function specifies propensities for events to occur, and the
idea
  of Wigner (or von  Neumann) that these events are  associated with
experiential
  qualities,  in  some very  generic  sense, but  allow  events to  occur in
both
  inanimate  and  animate systems.  However, I  focus  first on  those
particular
  events that are identifiable with  human conscious events, since we have
direct
  information about these.
  \newpage
  \noindent {\bf 5. Choice and Consciousness}

  William James concludes from a study ``of the particulars of the distribution
  of consciousness'' (as contrasted with our perhaps misleading intuition) that
  ``consciousness is at all times primarily a selecting agency''. He says also:
  ``It is to my mind utterly inconceivable that consciousness should have
nothing
  to do with a business to which it so fathfully attends''.

  But why should he, or anyone else,  even imagine that consciousness has
nothing
  to do with the  choices we  make? The reason,  of course, is  that this is
what
  classical physics tells us.

  Let me explain. The infant learns, early on, through concordance of
impressions
  gleaned from the five senses, including reports of others, to think that
things
  like apples  and  toys, etc. continue  to exist even when  no one is sensing
or
  actively thinking about them.  Classical physics extends this idea of
objective
  existence to the whole  world of inanimate objects:  all such things, large
and
  small, are  conceived to be  mere aggregates  of simple  localizable
properties
  that evolve according to local  deterministic laws. Functional structures,
such
  as  pistons  and  drive  shafts,  though   usefully  conceived  by us  as
whole
  functional   entities,  are  considered to  be   fundamentally  nothing but
the
  aggregates of the  interacting local parts of which  they are formed.
According
  to  classical  thinking, no  extra  property not   explainable in  terms of
the
  aggregrate of  simple  localized properties is  needed to  explain, at least
in
  principle, the  behavior of even the  most complex  physical structure. This
is
  the basic idea of  classical physics.  If we extend that  idea to the bodies
of
  human  beings  then  their   behaviors  should,  in  principle,  be
completely
  explainable in terms of their localizable components. Conscious thoughts do
not
  appear  in the   classical-physics  description,  and  hence, in  principle,
no
  reference  to such  things  should be needed  to  explain human  behaviour.
Any
  notion that  certain functional  features or aspects of  brain dynamics have
an
  experiential ``beingness as a whole''  that goes beyond the elemental
beingness
  of the   interacting  local  properties is  alien  to  classical  thinking,
and
  directly contradicts it if any dynamical role is given to such entities that
is
  not  completely  reducible  to the  local  dynamics of the  local  parts.
Thus,
  according  to the  ideas of  classical  mechanics, our  conscious  thoughts
are
  excess baggage: all physical behavior  would be just the same if the
functional
  structure of  the brain were  just what it is,  but no  conscious thoughts
were
  present.

  It is difficult to believe that  thoughts do nothing: that they are pure
excess
  baggage.  Yet how  is one to  make  sense of the   alternative idea  that
they,
  themselves, do  something that their  parts are not doing?  How can a
``whole''
  have an effect  that is not ultimately  just the effect of  its parts acting
in
  unison?

  Our point  of departure  is the fact  that in  (Heisenberg-von
Neumann-Wigner)
  quantum field theory there are two  distinct dynamical processes. They are
most
  clearly displayed in the so-called Heisenberg picture, or representation.
There
  the local operators of the theory  evolve according to the Heisenberg
equations
  of motion, which  are the  Heisenberg-picture counterparts  of the
Schroedinger
  equation.  These   equations  generate  from the  operators  located  along
any
  spacelike  surface  (or  constant-time  slice) the  operators at  all
spacetime
  points, i.e.,  at all points,  from the  infinite past to the  infinite
future.
  This is analogous to the situation in  classical mechanics, where the
classical
  equations of motion generate, from values on one space-like surface, the
values
  of all quantities at all times and places. But this part of the dynamics is,
in
  the quantum  case, only half  the story, and  the relatively  trivial
kinematic
  part at that. The nontrivial part of the dynamics is the part that controls
the
  evolution of the (Heisenberg picture) state of the universe. This part
consists
  of selections that are not determined by the local deterministic aspects of
the
  quantum  dynamics.  Orthodox  quantum  theory  says that  these  selections
are
  determined by pure  chance, but the  simplest naturalistic  possibility is
that
  they are  controlled by some nonlocal  aspect of the  physical universe. If,
in
  the case of  brain process,  this aspect can  be identified  with our
conscious
  thoughts, then consciousness would be a bona fide selecting agency. Because
the
  selection events are events they do not have separate parts: each quantum
event
  is a   selection and   actualization,  all at  once,  of a  spatially
extended
  structure of propensities.

  How could such a quantum process of selection and actualization work?

  \noindent {\bf 6. General Description of Brain/Mind Dynamics}

  Before  going  into the   mathematical  details of  the  model, I  give a
brief
  general description  of my  conception  of how the quantum brain/mind  works.
  For a more detailed description see reference [4].

  Each conscious event is the felt event  that actualizes a certain
``executive''
  pattern of brain  activity. This  pattern endures long  enough for it to
become
  ``facilitated'':  i.e.,  to become  etched into the  physical  structure of
the
  neurons in such a way that a subsequent excitation of part of the pattern
tends
  to spread to the whole  pattern. The sequence of  conscious thoughts
associated
  with a  given brain  is  represented by a  sequence of   actualizations of
such
  patterns. The patterns in such a sequences have, in general, a large
carry-over
  of  components from one  pattern to  the next. Thus  the  sequence of
executive
  patterns has the structure of a  ``marching band'' that marches into and out
of
  existence,  with new  parts  coming into  being at each  step, and  older
parts
  fading  out.  The  ``feel'' of  each  thought   expresses the   intentional
and
  attentional   content of  the  associated   actualized  executive  pattern.
The
  pervading experience of  the presence of an  enduring ``I'' is the felt
process
  of  continually   re-actualizing the  slowly  changing  peripheral  part of
the
  executive   pattern.  This  part  provides the   over-all  orientation  for
the
  executively  controlled part  of the  mind/brain process. The  sequence of
felt
  events that  actualize the  executive patterns  constitutes a  tiny part of
the
  brain activity: it rides on a vast substrate of unconscious brain activity
that
  is controlled by the  local deterministic process  governed by the equations
of
  local quantum  field theory. Each  executive pattern consists of a template
for
  action that is constructed largely from components of earlier templates, and
it
  issues its  directives to  the  lower-level  processes simply by  the
automatic
  spreading  of  the  neural   patterns of   excitations  that  comprise  it.
The
  processing is analog, not digital,  with a continual inflow of information
from
  the  environment,  to which  the body  and brain  adapt.  Although  the
analog
  process can be  simulated, at great  expense, by a digital  computer, the
issue
  here  pertains to how  real  brain tissues  and aqueous  ionic  solutions,
etc.
  function in real time.

  Due to the quantum nature of the brain, and in particular to point a)
mentioned
  in section 3  above, the brain state  must evolve, via the  local
deterministic
  process  determined  by the  equations of  local  quantum field  theory, into
a
  superposition   of states  each of  which  contains  at the   executive level
a
  different   alternative  possible   template for  action.  Each  alternative
is
  represented,  during some  brief time  interval, by a relative  stable
enduring
  pattern  of neural   activity, and  this  stability  constitutes the
condition
  required  for an  event to  occur. The  ``second   process'' now  enters. It
is
  represented in the  physical realm  (i.e., in Hilbert  space) by a selection
of
  one of these  alternative possible  states, each of which  specifies a
distinct
  template for action.

  According to orthodox quantum ideas, this selection event is controlled by
pure
  chance.  The use  of  ``pure  chance'' in  a  pragmatic  context is
completely
  acceptable. But it is not  acceptable at the level  of ontology. In the
context
  of a naturalistic science  some explanation in  terms of physical quantities
is
  needed, at  least in  principle, for how the  particular  reality that
actually
  appears is picked out from the collection of alternative possibilities that
are
  created by the local  deterministic part of the dynamics.

  The simplest  naturalistic possibility  is that the  selection is controlled
by
  the state  vector itself,  since this  vector, and  its changes,  represent
the
  physical  reality. A most  natural  possibility  would be for  the choice to
be
  controlled by  the aspect of  the state vector  that specifies  the
environment
  that defines the possible states between which the selection event must
choose.
  In our case that  aspect would be the  state of the brain  itself, or,
perhaps,
  even the aspect of the brain associated with the ``I'' mentioned above. In
this
  latter  case  it would  be the  ``I'',  as it  is  represented  in the
quantum
  dynamics, that selects  the sequence of templates  for action that controls
the
  behavior of the organism.

  But how could such a quantum process work?

  \noindent {\bf 7. Mathematical Formulation}

  My aim here to provide a mathematical model of causal quantum brain dynamics
in
  which the  quantum  selection process  is governed  by our  conscious
thoughts,
  rather than  by pure  chance; i.e.,  where the  notorious  stochastic
selection
  process of quantum  mechanics, called  the ``irrational''  element by Pauli,
is
  replaced by a causal  process in which our  conscious thoughts, acting as
whole
  entities not reducible to aggregates  of local properties, become the bona
fide
  selecting agents.

  Quantum electrodynamics  (extended to cover the  magnetic properties of
nuclei)
  is the theory that  controls, as far as we know,  the properties of the
tissues
  and the aqueous  (ionic) solutions  that constitute our  brains. This theory
is
  our paradigm basic physical theory,  and the one best understood by
physicists.
  It describes accurately,  as far as we know, the  huge range of actual
physical
  phenomema involving the materials encountered in daily life. It is also
related
  to classical electrodynamics in a particularly beautiful and useful way. I
take
  it as the basis of this work.

  In this section I assume the reader to have some knowledge of the principles
of
  quantum    electrodynamics,  and the   notations used  to  describe  it. I
draw
  particularly on references [14] and  [15], which describe in detail the
natural
  connection between quantum electrodynamics and classical electrodynamics.

  In the low-energy regime  of interest here it  should be sufficient to
consider
  just the  classical part of  the photon  interaction defined  in [14]. Then
the
  explicit expression for the unitary  operator that describes the evolution
from
  time $t_1$ to time $t_2$ of the quantum elecromagnetic field in the presence
of
  a set $L = \{L_i\}$ of specified  classical charged-particle trajectories,
with
  trajectory $L_i$ specified by the  function $x_i(t)$ and carrying charge
$e_i$,
  is
  $$U[L;t_2,t_1]=\exp<a^*\cdot J(L)>\exp<-J^*(L)\cdot a>
  \exp[-(J^*(L)\cdot J(L)/2)],
  $$
  where, for any $X$ and $Y$,
  $$
  <X\cdot Y>\equiv
  \int  d^4k  (2\pi)^{-4}2\pi \delta^+(k^2) X(k)\cdot Y(k),
  $$
  $$
  (X\cdot Y)\equiv \int d^4k(2\pi)^{-4}i(k^2+i\epsilon)^{-1} X(k)\cdot Y(k),
  $$
  and  $X\cdot Y =  X_{\mu}  Y^{\mu}  = X^{\mu}
  Y_{\mu}$.   Also,
  $$
  J_{\mu}(L;  k)\equiv  \sum_i   -ie_i\int_{L_i}  dx_{\mu}\exp(ikx).
  $$
  The integral along the trajectory $L_i$ is
  $$ \int_{L_i} dx_{\mu}\exp(ikx) \equiv
  \int_{t_1}^{t_2} dt (dx_{i\mu}(t)/dt) \exp(ikx).
  $$
  The $a^*(k)$ and $a(k)$ are the photon creation and annihilation operators:
  $$
  [a(k),a^*(k')] = (2\pi)^3  \delta^3 (k-k') 2k_0.
  $$

  The operator  $U[L; t_2, t_1]$  acting on the  photon vacuum  state creates
the
  coherent  photon  state that is the quantum-theoretic  analog of  the
classical
  electromagnetic field generated by  classical point particles moving on the
set
  of trajectories  $L=\{L_i\}$ between times $t_1$ and $t_2$.

  The $U[L; t_2, t_1]$ can  be decomposed into  commuting contributions from
  the various values of $k$. The general coherent state can be written
  $$
  |q,p>\equiv \exp i(<q \cdot P>- <p \cdot Q>)|0>,
  $$
  where $|0>$ is the photon vacuum state and
  $$
  Q(k) = (a_k + a^*_k)/\surd 2
  $$
  and
  $$
  P(k) = i(a_k - a^*_k)/\surd 2,
  $$
  and $q(k)$ and $p(k)$ are two functions defined (and square integrable)
  on the mass shell $k^2=0$, $k_0\geq 0$. The inner product of two coherent
  states is
  $$
  <q,p|q',p'>=\exp -(<q-q'\cdot q-q'>+<p-p'\cdot p-p'>+2i<p-p'\cdot q+q'>)/4.
  $$
  There is a decomposition of unity
  $$
  I = \prod d^4k(2\pi)^{-4}2\pi \delta ^+(k^2) \int dq_k dp_k/\pi
  $$
  $$
  \times \exp (iq_k P_k-ip_k Q_k)|0_k><0_k|\exp-(iq_k P_k-ip_k Q_k).
  $$
  Here meaning can be given by quantizing in a box, so that that the variable
  $k$ is discretized.
  Equivalently,
  $$
  I=\int d\mu (q,p) |q,p><q,p|,
  $$
  where $\mu(q,p)$ is the appropriate measure on the functions  q(k) and p(k).
  Then if the state $|\Psi><\Psi|$ were to jump to $|q,p><q,p|$ with
probability
  density $<q,p|\Psi><\Psi|q,p>$, the resulting mixture would be
  $$
  \int d\mu (q,p) |q,p><q,p|\Psi><\Psi|q,p><q,p|,
  $$
  whose trace is
  $$
  \int d\mu (q,p) <q,p|\Psi><\Psi|q,p> = <\Psi|\Psi>.
  $$

  To  represent the  limited  capacity of  consciousness  let us  assume, in
this
  model,  that  the  states of   consciousness   associated with  a  brain can
be
  expressed   in  terms  of a   relatively  small   subset of  the  modes  of
the
  electromagnetic field in the brain  cavity. Let us assume that events
occurring
  outside  the brain  are keeping  the state  of the  universe  outside the
brain
  cavity  in a   single  state,  so that  the  state  of the   brain can  also
be
  represented  by a  single state.  The brain  is  represented, in  the method
of
  Feynman, by a  superposition of the  trajectories of the  particles in it,
with
  each   element of  the    superposition   accompanied  by  the
coherent-state
  electromagnetic field that this set of trajectories generates. Let the state
of
  the electromagnetic field  restricted to the modes that represent
consciousness
  be called
  $|\Psi (t)>$.  Using the decomposition  of unity one can  write
  $$
  |\Psi (t)> =\int d\mu  (q,p)  |q,p><q,p|\Psi (t)>.
  $$
  Hence the state at time $t$ can be represented by the function
$<q,p|\Psi(t)>$,
  which is a complex-valued function over the set of arguments $\{ q_1, p_1,
q_2,
  p_2,  ...  , q_n, p_n  \}$,  where n is  the  number of  modes   associated
with
  $|\Psi>$.  Thus in  this  model the  contents of the consciousness associated
  with a brain is represented in  terms of this function  defined over a
  $2n-$dimensional space: the $i$th conscious event is represented by the
  transition
  $$
  |\Psi_i (t_{i+1})> \longrightarrow
|\Psi_{i+1}(t_{i+1})>=P_i|\Psi_i(t_{i+1})>,
  $$
  where $P_i$ is a projection operator.

  For each allowed  value of $k$ the  pair of numbers  $(q_k,p_k)$ represents
the
  state of motion of the  $k$th mode of the  electromagnetic field. Each of
these
  modes is defined by a particular wave pattern that extends over the whole
brain
  cavity. This pattern is an oscillating structure something like a sine wave
or
  a cosine wave. Each mode is fed by the  motions of all of the charged
particles
  in the brain. Thus each mode is a representation of a certain integrated
aspect
  of the activity of the  brain, and the collection of values $q_1,p_1,...,p_n$
  is a compact representation of certain aspects the over-all activity of the
  brain.

  The state $|q,p>$ represents the conjunction, or collection over the set of
all
  allowed  values of $k$,  of the  various states  $|q_k,p_k>$.  The function
  $$
  V(q,p,t)= <q,p|\Psi (t)><\Psi (t)|q,p>
  $$
  satisfies  $0\leq V(q,p,t)  \leq 1$, and it  represents,  according to
orthodox
  thinking, the  ``probability'' that a  system that is  represented by a
general
  state  $|\Psi (t)>$  just before  the time  $t$ will be  observed  to be in
the
  classically  describable state $|q,p>$  if the observation  occurs at time
$t$.
  The coherent states $|q,p>$ can, for various mathematical and physical
reasons,
  be   regarded as  the  ``most    classical'' of  the   possible  states  of
the
  electromagnetic quantum field.

  To  contruct a  causal  dynamics in  which the  state of   consciousness
itself
  controls the  selection of the next  state of  consciousness one must specify
a
  rule that determines, in terms of the  evolving state $|\Psi_i (t)>$ up to
time
  $t_{i+1}$, both  the time $t_{i+1}$  when the next  selection event occurs,
and
  the state    $|\Psi_{i+1}(t_{i+1})>$ that is  selected  and  actualized by
that
  event.

  In  the  absence of    interactions, and  under  certain  ideal   conditions
of
  confinement,  the  deterministic normal law of  evolution  entails that in
each
  mode $k$  there is an   independent rotation  in the  $(q_k,p_k)$  plane with
a
  characteristic  angular  velocity $\omega_k =  k_0$. Due to  the effects of
the
  motions of the  particles there will  be, added to this, a  flow of
probability
  that will tend to concentrate the probability in the neighborhoods of a
certain
  set of ``optimal'' classical states $|q,p>$. The reason is that the function
of
  brain  dynamics  is to  produce  some  single  template for  action,  and to
be
  effective this  template must be a  ``classical'' state,  because, according
to
  orthodox ideas,  only these can be  dynamically robust in  the room
temperature
  brain [16]. According to the  semi-classical description of the brain
dynamics,
  only one  of these   classical-type states  will be  present, but  according
to
  quantum  theory  there must  be a  superposition  of many  such
classical-type
  states,  unless  collapses  occurs at  lower  (i.e.,  microscopic)  levels.
The
  assumption here is that no collapses  occur at the lower brain levels: there
is
  absolutely no empirical evidence, or theoretical  reason, for the occurrence
of
  such lower-level brain events.

  So in this model the probability will  begin to concentrate around various
  locally optimal  coherent states,  and hence around the  various (generally)
  isolated    points  $(q,p)$  in  the     $2n-$dimensional  space  at  which
  the quantity
  $$
  V(q,p,t)=<q,p|\Psi_i (t)><\Psi_i  (t)|q,p>
  $$
  reaches   a  local   maximum.  Each of   these  points   $(q,p)$   represents
a
  ``locally-optimal solution'' (at time $t$) to the search problem: as far as
the
  myopic  local   mechanical  process  can see  the  state  $|q,p>$  specifies
an
  analog-computed ``best'' template for  action in the circumstances in which
the
  organism  finds  itself. This  action can be  either  intentional  (it tends
to
  create in the future a certain state  of the body/brain/environment complex)
or
  attentional  (it  tends to  gather   information), and the  latter  action is
a
  special case of the former. As discussed in [4], the intentional and
  attentional character of these actions is a consequence of the fact that the
  template for action actualized by the quantum brain event is represented as
  a projected body-world schema, i.e., as the brains projected representation
  of the body that it is controlling and the environment in which it is
situated.

  Let a  certain time  $t_{i+1}> t_i$ be  defined by  an (urgency)  energy
factor
  $E(t)=  \hbar(t_{i+1}- t_i)^{-1}$. Let  the value of  $(q,p)$ at the largest
of
  the local-maxima of $V(q,p,t_{i+1})$ be called
$(q(t_{i+1}),p(t_{i+1}))_{max}$.
  Then the simplest  possible  reasonable  selection rule  would be  given by
the
  formula
  $$
  P_i= |(q(t_{i+1}),p(t_{i+1}))_{max}><(q(t_{i+1}),p(t_{i+1}))_{max}|,
  $$
  which entails   that
  $$
   |\Psi_{i+1}><\Psi_{i+1}|/      <\Psi_{i+1}|\Psi_{i+1}>=
  |(q(t_{i+1}),p(t_{i+1}))_{max}><(q(t_{i+1}),p(t_{i+1}))_{max}|.
  $$

  This rule could produce a tremendous speed up of the search process. Instead
of
  waiting until all  the probability  gets concentrated in  one state $|q,p>$,
or
  into a set of isolated  states $|q_i,p_i>$ [or  choosing the state randomly,
in
  accordance with  the probability  function  $V(q,p,t_{i+1})$, which could
often
  lead to a  disastrous result],  this simplest  selection  process would pick
the
  state $|q,p>$  with the  largest value of  $V(q,p,t)$ at the  time
$t=t_{i+1}$.
  This process does  not involve the  complex notion of  picking a random
number,
  which is a physically impossible feat that is difficult even to define.

  One important feature of  this selection process is  that it involves the
state
  $\Psi (t)$ as  a whole: the  whole function  $V(q,p,t_{i+1})$  must be known
in
  order to determine  where its maximum  lies. This kind of  selection process
is
  not available  in the  semi-classical  ontology, in which only  one
classically
  describable state exists at the  macroscopic level. That is because this
single
  classically    describable   macro-state  state  (e.g., some  one  actual
state
  $|q,p,t_{i+1}>$)     contains no   information  about  what the
probabilities
  associated either with itself or with the other alternative possibilities
would
  have been if the  collapse had not  occurred earlier, at  some micro-level,
and
  reduced the  earlier  state to some  single  classically  describable state,
in
  which, for  example, the  action potential  along each nerve  is specified by
a
  well  defined   classically  describable   electromagnetic  field.  There is
no
  rational  reason in quantum  mechanics for  such a micro-level  event to
occur.
  Indeed,  the  only  reason  to  postulate  the   occurrence of  such
premature
  reductions  is to assuage  the  classical intuition  that the
action-potential
  pulse along each  nerve ``ought to be  classically  describable even when it
is
  not  observed'',  instead of  being  controlled, when unobserved, by  the
local
  deterministic   equations of  quantum field  theory.  But the  validity of
this
  classical intuition is questionable if  it severely curtails the ability of
the
  brain to function optimally.

  A second  important feature of  this selection  process is  that the
actualized
  state   $\Psi_{i+1}$ is  the state  of the entire aspect of  the  brain that
is
  connected to  consciousness.  So the feel of  the conscious  event will
involve
  that aspect of the brain, taken as a whole.  The ``I'' part of  the state
$\Psi
  (t)$ is its slowly changing part. This  part is being continually
re-actualized
  by the sequence of  events, and hence specifies the  slowly changing
background
  part of the felt  experience. It is  this persisting  stable background part
of
  the sequence of  templates for action  that is providing  the over-all
guidance
  for the entire  sequence of selection  events that is  controlling the
on-going
  brain process itself.

  A  somewhat more  sophisticated  search  procedure  would be to  find the
state
  $|(q,p)_{max}>$, as before, but to identify it as merely a candidate that is
to
  be examined  for its  concordance with the  objectives  imbedded in the
current
  template. This is what a good search  procedure ought to do: first pick out
the
  top  candidate  by  means of a   mechanical  process,  but  then  evaluate
this
  candidate by a  more refined  procedure that could block  its acceptance if
it
  does not meet specified criteria.

  It may at  first seem strange  to imagine that  nature could  operate in such
a
  sophisticated  way. But it  must be remembered  that the  generation of a
truly
  random   sequence is   itself a  very   sophisticated  (and  indeed
physically
  impossible)  process, and that what  the physical sciences  have understood,
so
  far, is only the  mechanical part of nature's  two-part process. Here it is
the
  not-well-understood   selection  process  that is under   consideration. I
have
  imposed on this  attempt to understand  the selection  process the
naturalistic
  requirement that the whole process be  expressible in natural terms, i.e.,
that
  the   universal   process be  a  causal    self-controlling   evolution  of
the
  Hilbert-space   state vector  in which  all  aspects of  nature,  including
our
  conscious experiences, are efficacious.

  No attempt is made here to show that the quantum statistical laws will hold
for
  the aspects of the brain's internal  dynamics controlled by conscious
thoughts.
  No such result has been  empirically verified. The  validity of the
statistical
  laws for  events in the  inanimate  world is  regarded as a  consequence of
our
  ignorance   of  the  actual   causes, and  of   certain a  priori
probability
  distributions. This is discussed in section 9.

  It  may  be   useful  to   describe  the   main   features  of  this   model
in
  simple terms.  If we imagine the brain to  be, for example, a uniform
  rectangular  box  then each  mode $k$  would  correspond  to wave  form that
is
  periodic  in all  three  directions:  it would  be formed  as a  combination
of
  products of sine waves  and cosine waves, and would  cover the whole
box-shaped
  brain. (More realistic conditions are needed, but this is a simple
proto-type.)
  Classically there  would be an  amplitude for this wave,  and in the absence
of
  interactions with the  charged particles this  amplitude would undergo a
simple
  periodic motion in time. In analogy  with the coordinate and momentum
variables
  of an oscillating pendulum there are two variables, $q_k$ and $p_k$, that
describe
  the motion of the  amplitude of the  mode $k$. With a  proper choice of
scales for
  the variables  $q_k$ and $p_k$  the motion of  the amplitude  of mode $k$ if
it
  were  not   coupled  to  the   charges  would  be a   circular   motion  in
the
  $(q_k,p_k)$-plane.  The  classical theory  would say that the  physical
system,
  mode $k$,  would be  represented  by a point  in  $q_k,p_k$ space.  But
quantum
  theory says that  the physical system,  mode $k$, must be  represnted by a
wave
  (i.e., by a wave function) in $(q_k,p_k)$  space. The  reason is  that
interference
  effects   between the  values  of this  wave   (function) at   different
points
  $(q_k,p_k)$ can be exhibited, and therefore it is not possible to say  the
full reality is
  represented by any single value of $(q_k,p_k)$: one must acknowledge the
reality
  of the whole wave. It is possible to associate  something like a
``probability
  density''  with this  wave, but the corresponding  probability  cannot be
concentrated at a
  point: in units  where Planck's  constant is unity the  bulk of the
probability
  cannot be  squeezed into a  region of the  $(q_k,p_k)$ plane  of area less
that
  unity.

  The  mode  $k$  has  certain   natural  states   called   ``coherent
states'',
  $|q_k,p_k>$.  Each  of these  is  represented in   $(q_k,p_k)$-space  by a
wave
  function that has a ``probability density'' that falls off exponentially as
one
  moves in  any  direction away  from the  centerpoint  $(q_k,p_k)$  at which
the
  probability  density is  maximum.  These coherent  states are in  many ways
the
  ``most classical'' wave functions allowed by quantum theory [17], and a
central
  idea of  the present  model is  to specify  that it is  to one of  these
``most
  classical''  states that the  mode-$k$  component of the  electromagnetic
field
  will  jump, or   collapse,  when an   observation  occurs.  This
specification
  represents  a  certain  ``maximal''  principle:  the  second  process, which
is
  supposed to  pick out and  actualize some  classically  describable reality,
is
  required  to pick out  and  actualize one of  these  ``most  classical'' of
the
  quantum states. If this selection/actualization process really exists in
nature
  then the  classically  describable states that  are actualized  by this
process
  should  be  ``natural  classical  states''  from  some point  of  view.
The
  coherent  states  satisfy this  requirement.  This  strong,  specific
postulate
  should be  easier to  disprove,  if it is  incorrect,  than a  vague or
loosely
  defined one.

  If we  consider a  system  consisting of a  collection  of modes  $k$, then
the
  generalization of the single coherent state $|q_k,p_k>$ is the product of
these
  states, $|q,p>$.  Classically this system would be  described by specifying
the
  values all of the classical variables $q_k$ and $p_k$ as functions of time.
But
  the  ``best'' that  can be  done  quantum  mechanically is  to  specify that
at
  certain  times  $t_i$ the  system is  in one of  the  coherent  states
$|q,p>$.
  However,   the   equations  of  local   quantum  field   theory  (here
quantum
  electrodynamics)  entail  that if the  system  starts in such a  state then
the
  system will,  if no  ``observation'' occurs,  soon evolve into  a
superposition
  (i.e., a linear combination) of many  such states. But the next
``observation''
  will then reduce it again to some classically describable state. In the
present
  model each a human  observation is identified as a  human conscious
experience.
  Indeed,   these are  the  same   observations  that the   pragmatic
Copenhagen
  interpretation of Bohr  refers to, basically. The `happening' in a  human
brain
  that corresponds to such an observation is, according to the present model,
the
  selection and actualization of the corresponding coherent state $|q,p>$.

  The quantity $V(q,p, t_{i+1})$ defined  above is, according to orthodox
quantum
  theory, the  predicted  probability  that a system  that is in  the state
$\Psi
  (t_{i+1})$ at  time $t_{i+1}$  will be  observed to be in  state $|q,p>$ if
the
  observation   occurs at  time  $t_{i+1}$.  In the  present  model the
function
  $V(q,p,t_{i+1})$  is  used to  specify not a   fundamentally  stochastic
(i.e.,
  random or   chance-controlled)  process but  rather the  causal  process of
the
  selection and  actualization of some particular  state $|q,p>$. And this
causal
  process is  controlled by features of  the quantum brain  that are specified
by
  the Hilbert space representation of  the conscious process itself. This
process
  is a  nonlocal process  that rides on  the local  brain process,  and it is
the
  nonlocal selection process that, according to the principles of quantum
theory,
  is required to enter whenever an observation occurs.
  \newpage
  \noindent {\bf 8. Qualia: The Feel Of An Actualization}

  According to the theory described  here, a human conscious event is the
reality
  that is represented in  the model as a felt event  that actualizes an
executive
  template for  action in a  human  brain, and the  flow of  consciousness is
the
  reality that  is represented  as the sequence  of felt events  that actualize
a
  sequence of  executive  templates for  action in a  human brain.  The
conscious
  ``I''   is  the   reality  that  is    represented as  the    sequence of
felt
  re-actualizations   of  the  slowly  changing   background  structure  in
these
  templates   for  action.  This  background   structure   provides the
over-all
  orientation for the ongoing mind/brain process. Since the whole quantum
process
  takes place in the realm of  potentialities, or probabilities, or
propensities,
  which are  mind-like in  character, and these  quantities  pertain only to
felt
  events,   which  are  just  the    actualizations  of  other
potentialities,
  probabilities, and propensities, the  whole quantum ontology has an
essentially
  mind-like character:  ontologically speaking,  everything is mind like. Yet
all
  of  these   mind-like  things  are   represented    mathematically in  terms
of
  Hilbert-space   vectors,  which is what represents, in  quantum  mechanics,
the
  physical  aspect of  nature.  Thus this model integrates into one
mathematical
  structure  the  mental and  physical  aspects of  nature.  The  conflation of
  mind and matter by quantum theory was, of  course, a feature well appreciated
  its founders.

  \noindent {\bf 9. Quantum Statistics}

  If the process of selection and actualization of ``the actual'' in human
brains
  is governed by a nonlocal causal  process, rather than by pure chance, then
one
  must  naturally  expect  analogous  causal  processes  to be  occurring
  elsewhere in  nature. If we assume  that the selection  process is in all
cases
  controlled by a  causal process then  it must be explained  why the
statistical
  rules of  quantum theory hold  in those cases  where they have  been tested
and
  validated.

  An explanation can be constructed as follows. Consider an n-dimensional
Hilbert
  space of points $(z_1,z_2, ...,z_n)$, where, each for each $i$,
$$z_i=x_i+iy_i=
  r_i \exp i\theta_i$$ is a  complex number, and  $r_i\geq 0 $. This space can
be
  imbedded  in a   2n-dimensional  real  space of  points  $(x_1, y_1,  x_2,
y_2,
  ...,x_n, y_n)$, and each unitary  transformation in the Hilbert space
generates
  an orthogonal  transformation in the  real space. The  volume in the real
space
  defined by the  intersection of the  unit ball centered at  the origin with
the
  collection of rays from  the origin that pass  through a region $R$ on the
unit
  sphere is invariant under  any orthogonal  transformation, and hence also
under
  the image in  real space of  any unitary  transformation in  the Hilbert
space.
  Thus the  volume (=surface  area) of any  region $R$ of the  unit sphere in
the
  real space is  invariant under the  image of any unitary  transformation in
the
  Hilbert space.

  Since  dynamical  evolution, and most  symmetry  operations in  the the
Hilbert
  space, are  generated  by  unitary  transformations,  the a priori
probability
  density of  unit vectors  in Hilbert  space should  be invariant  under
unitary
  transformations.   Thus it is  reasonable to  assign to  any  region $R$ on
the
  surface of the  real unit  sphere an a priori  probability  equal to the
volume
  (=surface area) of that region $R$.

  This a priori probability rule can be  used in the following way. Suppose
that,
  as in  our brain  case,  there is,  for a  given  state  $\Psi_i$, a  rule
that
  specifies a candidate  projection operator $P_i$,  and that if the passage
from
  state $\Psi_i$ to state $P_i\Psi_i$ is  not  ``blocked''  then  the
transition
  proceeds. If $P_i=I$, where I is the identity operator, then the passage is
not
  blocked, since a  change into itself  is no change at all,  and if $P_i=0$
then
  the  passage must  be  blocked, since  a  transition to  the null  state is
not
  allowed.

  But then what is the rule that determines whether the passage is blocked?

  According to the idea behind the present theory everything that enters into
the
  dynamics  is  represented in  Hilbert space:  nothing  dynamically
significant
  stands outside  the Hilbert  space of the  universe! And the  dynamics is to
be
  specified in  terms of the  state of the  universe, or perhaps  in terms of
the
  full history of states $$(...,\Psi_{i-2}, \Psi_{i-1}, \Psi_i).$$

  The   simplest form for the ``blocking  rule'' is that the states $\Psi_i$
and
  $P_i\Psi_i$  determine a  state $\Phi$  of unit  norm that lies  in the
complex
  2-dimensional  subspace  generated by  $\Psi_i$ and   $P_i\Psi_i$, and that
the
  transition from the state $\Psi_i$ to the state $P_i\Psi_i$ proceeds unless
for
  some representative of the state  $\Psi_i$, which is defined only up to a
phase
  factor, the  direct path from  $\Psi_i$ to  some  representative of
$P_i\Psi_i$
  intersects the ray $\Phi$

  The geometric situation  is this. The state  $\Psi_i$ can be represented in
the
  2-dimensional    Hilbert space   generated by   $\Psi_i$ and $P_i\Psi_i$ by
the
  continuum of pairs of complex numbers $$(z_1, z_2)=(\exp i\phi , 0); 0\leq
\phi
  \leq1,$$ and the state $P_i\Psi_i$ can then be  represented by the continuum
of
  pairs  $$(\cos^2  \theta \exp  i\phi, \sin  \theta\cos  \theta  \exp i\phi
\exp
  i\chi)$$ with  $0\leq \phi \leq 2\pi$  and $0\leq \chi  \leq 2\pi$. The
overall
  phase  factor $\exp  i\phi$  drops out of  all  computations and  can be set
to
  unity. The phase factor $\chi$ reflects an arbitrary choice of the phase of
the
  basis vector associated with the  component $z_2$, and it is assumed that
there
  is a representative of $P_i\Psi_i$ for each value of  $\chi$. The ``direct
path''
  from a   representative of  $\Psi_i$ to a   representative of  $P_i\Psi_i$
can be
  traced out by  allowing the  value of  $\theta$ to run  from zero to its
actual
  value. Allowing  $\theta$ to run from  zero to $\pi /2$  and $\chi$ to run
from
  zero to $2  \pi$  generates  a   2-dimensional  spherical surface  $S_{1/2}$
of
  radius $1/2$  centered at  $z_1 =  1/2$. The vectors  $\Phi$ are defined as
the
  set of unit-normed  vectors from the  origin  $z_1=z_2=0$, or as the
equivalent
  parallel vectors of norm $1/2$ from the  center of $S_{1/2}$. A uniform
  distribution   of the   unit-normed  vectors  $\Phi$  on the  unit  2-sphere
is
  equivalent  to a  uniform   distribution  of points  on the   spherical
surface
  $S_{1/2}$.            Notice    that   a     point
  $$(\cos^2\theta', 0,\sin\theta'\cos\theta'\cos\chi',
                          \sin\theta'\cos\theta'\sin\chi')$$
  on $S_{1/2}$ blocks some direct path in $S_{1/2}$ from the representative
  $(1, 0, 0, 0)$ of $\Psi_i$ to some representative of $P_i\Psi_i$ if and only
if
  $\theta'$ satisfies $0\leq \theta' \leq \theta$

  In some situations, namely those in  which the realities that are governing
the
  second process  are human  conscious  experiences, we have  direct knowledge
of
  what the governing  realities are:  they are exactly the  conscious
experiences
  that are controlling the second process. But in cases where the collapse of
the
  wave function is associated with, say, an event in a Geiger counter, we are
not
  privy to the form of the controlling  realities. So in these cases we must
fall
  back to  statistical  considerations.  According to the model  described
above,
  there is a  vector  $\Phi$ that  determines  whether or  not the  collapse
will
  occur,  but we  are  ignorant  of what  it is.  But  the a  priori
probability
  distribution  for the location  of the vector  $\Phi$  corresponds to a
uniform
  distribution  over the  spherical surface  $S_{1/2}$. The  probability that
the
  transition from  $\Psi_i$ to  $P_i\Psi_i$ will be blocked  is then equal to
the
  fraction of  the surface area  of $S_{1/2}$  that is covered  as $\theta'$
runs
  from zero to $\theta$. This probability is $1-\cos^2\theta$. Hence the a
priori
  probability that the transition will  occur is $\cos^2\theta$. This is the
same
  as   $|P_i\Psi_i|^2/|\Psi_i|^2$, which is what  quantum theory  predicts. So
in
  this model the  statistical  predictions of  quantum theory  would arise from
a
  combination  of our  ignorance  of the true  causes,  with an a  priori
uniform
  probability  distribution over an  appropriate 2-sphere of  the real image of
a
  Hilbert  space  vector  $\Phi$  that  determines  whether  the  transition to
a
  specified state occurs or not.

  \noindent {\bf 10. Remarks }

  1. Quantum brain theory  has been characterized as  ``A solution in search of
a
     problem''. A first question, in this connection, is whether a
semi-classical
     model of the  brain---e.g., a model  in which the action  potential on
every
     neuron is regarded as a well-defined classically describable
electromagnetic
     pulse---is capable  of generating solutions to  search problems as
     quickly as the brain  actually does it, or  whether a quantum mechanism
such
     as the  hydrodynamic  effect,  or the  picking of the  most  likely
solution
     mentioned above, is needed. The way  in which a classical brain could
search
     for suitable  templates for action  (or recognize  patterns) is not known
at
     present in  enough  detail to make an  estimate  of the  claasically
allowed
     rapidities possible . But it seems  reasonable that nature would make use
of
     the quantum possibilities for speeding up the search processes.

  2. This question  of speed is,  however, not  the only  relevant
consideration.
     Even if a  semi-classical  model  were fast  enough the question would
arise
     why a  dynamically  inert  psychical  element is  present at all  in
nature.
     Wigner emphasized that in the rest of physics every action of one thing
upon
     another is  accompanied  by a  reaction of the  second back on  the first.
A
     dynamically  inert psychic  reality could  have no survival  value, hence
no
     physical  reason to exist.  Yet it seems  absurd to think  that something
so
     different from its supposedly classical physical foundation could arise
just
     by accident.

  3. The  model   described  here is   heretical in   attempting  to  replace
the
     irreducible element of chance in quantum theory by a nonlocal causal
process
     in Hilbert space.  Indeed, in my  earlier works on the  subject I adhered
to
     the  orthodox idea  that the  statistical   predictions are  inviolate.
Even
     adhering  to that  stricture,  the evident   mentalistic  character of
basic
     elements in  quantum mechanics (i.e.,  the existence of  nothing material
or
     substantive; but  merely  probabilities, nonlocal  selection events, and
the
     experiences of observers) suggested  that the experiential aspects of
nature
     were  closely  tied to  its   fundamentally  quantum  nature. But  if
mental
     entities  are really  entering into  the ontology  in a basic  way, it
seems
     unnatural  not  to give  them a  genuine   dynamical role,  rather  than
the
     illusory one that they  would have if an  irreducible element of chance
were
     really controlling the selection process. In any case, perhaps this
spelling
     out of a  simple mathematical  model may  convey better  than words the
fact
     that quantum  theory naturally  accomodates a conception  of nature in
which
     there is, in the human  brain, a nonlocal  physical process of selection
and
     actualization  that:  (1),  supervenes  over the  local process  that is
the
     quantum  anologue of the  local  process of  classical  physics; (2), is
not
     reducible to any local process; and (3), plays a bona fide executive role
in
     the determination of our mental and physical actions.

  4. The events in this second process have an ontological character that
differs
     greatly  from that  of the  local process:  the  events abruptly  select
and
     actualize, via a global process, new  states of the physical system,
whereas
     the  local  process   merely  evolves in  a  continous   mechanical  way
the
     potentialities  for these  actual events. It  is therefore  natural that
the
     events should be  endowed with a  different kind of  beingness: i.e., with
a
     certain ``actualness''  that goes beyond the mere  ``tendency'' character
of
     what is generated by the  local process. Since  this actualization event
is,
     in the  case of  brain events,   simultaneously both  an  actualization of
a
     template for  action and an  implanting of the form of  this template into
a
     memory  structure, in the  form of its  projected  functional effects on
the
     body and  its  enviroment, it is not  unnatural  that the  beingness of
this
     brain event should be an embodiment or  representation of the functional
     character of this event.

  \noindent {\bf References}

  1. N. Bohr, See ref.4 p. 63/64.

  2. D. Bedford and H.P. Stapp, Synthese 102, 139-164, 1995; H.P. Stapp,
     Phys. Rev. A49, 4257, 1994; Ref.4, p.6.

  3. R.P. Feynman, Feynman Lectures, Chapter 21.

  4. H.P.Stapp, Mind, Matter, and Quantum Mechanics, Spinger-Verlag,
Heidelberg,
     1993. Chapter 6.

  5. A. Einstein, in A. Einstein: Philosopher-Scientist, ed. P.A. Schilpp,
Tudor,
     New York, 1951. p.667-673.

  6. N. Bohr, See ref. 4, Chapter 3

  7. D. Bohm and B.J. Hiley, The Undivided Universe: An Ontological
     Interpretation of Quantum Theory, Routledge, London, New York, 1993.

  8. H. Everett III, Rev. Mod. Phys. 29, 463, (1957).

  9. W. Heisenberg, Physics and Philosophy, Harper Row, New York, 1958,
     Chapter III.
  10. E. Wigner, in The Scientist Speculates, ed.I.J. Good, Basic Books,
      New York, 1962.

  11. G.C.Ghirardi, A. Rimini, and Weber, Phys. Rev. D34, 470 (1986);
      G.C.Ghirardi, P.Pearle, and A.Rimini, A42, 78 (1990)

  12. M. Gell-Mann and J. B. Hartle, in Proc. 3rd Int. Sympos. on Quantum
      Mechanics in the Light of New Technology, eds. S. Kobayashi, H. Ezawa,
      Y. Murayama, and S. Nomura, Phys. Soc. of Japan.

  13. R. Omnes, The Interpretation of Quantum Mechanics, Princeton Univ. Press,
      Princeton NJ, 1994.

  14. H.P.Stapp, Phys. Rev. D28, 1386 (1983)

  15. T. Kawai and H.P. Stapp, Phys. Rev. D52, 2484-2532, (1995)

  16. H.P. Stapp, in Symposium on the Foundations of Modern Physics 1990,
  P.Lahti and P. Mittelstaedt eds., World Scientific, Singapore. Sec. 3;
  and in ref.[4] p. 130.

  17. R.J. Glauber, in Quantum Optics, S.M. Kay and A. Maitland, eds.
  Academic Press, London and New York, 1970; T.~W.~B. Kibble in ibid;\\
  H.P. Stapp, in Quantum Implications: Essays in Honour of David Bohm,
  B.J. Hiley and F.David Peats eds., Routledge and Paul Kegan Ltd.,
  London and New York, 1987.
  \end{document}